\documentclass[12pt]{article}
\usepackage[T1]{fontenc}
\usepackage{times}
\usepackage[margin=1in]{geometry}
\usepackage{setspace}
\usepackage{amsmath,amssymb}
\usepackage{graphicx}
\usepackage{url}
\usepackage{hyperref}
\usepackage{tabularx}
\usepackage{booktabs}
\usepackage{float}
\usepackage{csquotes}
\usepackage{subcaption}
\usepackage{caption}
\usepackage{adjustbox}
\usepackage{tikz}
\usetikzlibrary{positioning, arrows.meta, shapes.geometric}
\usepackage{xcolor}

\usepackage{titlesec}
\titleformat{\section}{\centering\bfseries}{}{0pt}{}

\titleformat{\subsection}{\bfseries\leftskip0pt}{}{0pt}{}

\titleformat{\subsubsection}{\bfseries\itshape\leftskip0pt}{}{0pt}{}

\usepackage[utf8]{inputenc}
\usepackage[natbibapa]{apacite}
\usepackage{authblk} 

\title{A race to belief: How Evidence Accumulation shapes trust in AI and Human informants}

\author[1]{Johan Sebastián Galindez-Acosta}
\author[2]{Juan José Giraldo-Huertas\thanks{Corresponding author: juan.giraldo9@unisabana.edu.co}}

\affil[1]{University of La Sabana, Campus del Puente del Común, Km. 7, Autopista Norte de Bogotá, Chía, Cundinamarca 250001, Colombia}
\affil[2]{University of La Sabana, Campus del Puente del Común, Km. 7, Autopista Norte de Bogotá, Chía, Cundinamarca 250001, Colombia}

\date{} 

\begin{document}

\maketitle

\section*{Author Contributions}
Johan Sebastián Galindez-Acosta: Conceptualization, Data curation, Formal analysis, Investigation, Methodology, Project administration, Software, Validation, Visualization, Writing -- original draft, Writing -- review and editing.\\
Juan José Giraldo-Huertas: Data curation, Formal analysis, Investigation, Methodology, Supervision, Validation, Writing -- review and editing.

\section*{Acknowledgements}
We would like to express our gratitude to the Faculty of Psychology at the University of La Sabana for providing access to the student data utilized in this study. 

\section*{Declaration of Interest}
The authors declare that there is no conflict of interest regarding the publication of this article.

\maketitle
\begin{abstract}
Integration of artificial intelligence into everyday decision-making has created new dynamics in selective trust, yet the cognitive mechanisms underlying context-dependent preferences for AI versus human informants remain poorly understood. 

Bayesian Hierarchical Sequential Sampling Model (HSSM) was applied to examine how 102 Colombian university students navigate trust decisions across 30 hypothetical scenarios spanning epistemic (factual) and social (interpersonal) domains. Results demonstrate that context-dependent trust preferences are primarily driven by differences in drift rate (v), rate of evidence accumulation, rather than initial biases (z) or response caution (a). In epistemic contexts, participants exhibited strong negative drift rates (mean v = -1.26), indicating rapid evidence accumulation favoring AI, whereas social contexts yielded positive drift rates (mean v = 0.70) favoring human informants. Starting point parameters remained near-neutral (z = 0.52), suggesting minimal a priori bias.

Drift rate showed a strong and consistent within-subject relationship with signed confidence (Fisher-z-averaged r = .736, 95\% bootstrap CI [.699, .766]; 97.8\% of individual correlations positive, N = 93), demonstrating that model-derived evidence accumulation rate closely reflects participants’ moment-to-moment confidence. Findings may contribute to the perceived brittleness of trust in AI, which is known to collapse rapidly after errors, low-vigilance evidence processing in epistemic domains, where users defer trust without sufficient critical scrutiny. 

We interpret these patterns through the lens of epistemic vigilance theory, suggesting that contextually gated vigilance mechanisms modulate evidence accumulation differentially across domains. The study has implications for AI governance, highlighting the need for transparency mechanisms that maintain epistemic vigilance while preserving AI's efficiency advantages. By providing a mechanistic account of selective trust, this research advances theories of human-AI collaboration and offers empirical grounding for designing trustworthy intelligent systems aligned with human cognitive processes.
\end{abstract}

\section{Introduction}
The integration of artificial intelligence (AI) into daily life has forged a new epistemic landscape, fundamentally altering how individuals acquire knowledge and make decisions \citep{li2025reinforcement}. In this environment, generative agents like chatbots now compete directly with traditional human informants, like peers, family, and educators, for epistemic authority \citep{HuynhAichner2025, BuchananHickman2024}. This dynamic interaction necessitates a deeper understanding of selective trust: the cognitive mechanism for adaptively choosing whom and what to believe \citep{SchmidBleijlevensManiBehne2024, TongWangDanovitch2019}. While emerging evidence suggests that human trust in advice sources may be influenced by context, potentially leading to greater reliance on AI for objective, factual domains and on humans for subjective or personal matters \citep{osborne2025machine}, the cognitive mechanisms underlying these context-dependent preferences remain unclear. It is not yet known whether such trust shifts arise from deliberate, reflective evaluation or from rapid, automatic adjustments in evidence processing.

Research on selective trust in developmental science has established that even young children track an informant's accuracy to guide learning \citep{Mou2025, Yang2023}. Recent work with technological agents adds nuance: children may prioritize an accurate robot over an inaccurate human but often default to a social preference for people \citep{geng2025childrens}. Similar patterns, though developmentally modulated, emerge in adults, who exhibit greater epistemic trust in technological agents for factual tasks while retaining social preferences for humans in personal contexts \citep{li2023epistemic}; notably, younger children are more susceptible to human-like cues in machines, whereas older children tend to over-trust machines as informants \citep{hoehl, Stower2024}.

These findings are commonly interpreted through dual-process accounts, which propose that trust judgments arise from the interaction between a fast, intuitive System~1 and a slower, analytical System~2 responsible for evaluating reliability and transparency \citep{chiriatti2025system0, zhou2024mindlessly, okonkwo2025explainable}. Yet such models remain descriptive and offer limited insight into the computational mechanisms underlying context-dependent preferences for humans versus AI. Specifically, it remains unclear whether differences in trust reflect initial biases, varying levels of caution in responding, or differential rates of evidence accumulation during decision-making. Evidence further indicates that trust in humans and trust in AI are dissociable constructs, underscoring the need for mechanistic accounts capable of capturing these dynamics.

To advance this perspective, theoretical work highlights the role of epistemic vigilance, the suite of cognitive mechanisms humans use to assess the reliability of communicated information and protect against misinformation \citep{sperber2010epistemic}. In situations of selective trust, where individuals accept information without immediate verification based on past reliability or contextual cues, epistemic vigilance becomes central in calibrating trust \citep{levy2022trust}. These concepts have been extended by the Extended Epistemic Vigilance Framework (EEVF), which incorporates not only the evaluation of the source and the claim but also the characteristics of the receiver, including biases, socioemotional factors, and domain-specific sensitivities \citep{bielik2025developing}. Such frameworks suggest that trust in AI is likely gated by both contextual cues and self-reflective monitoring.

This theoretical gap is especially relevant for university students, who as digital natives engage extensively with AI, potentially refining their trust strategies in domain-specific ways \citep{abuzar2025university, chan2023students}. Adults, including students, display context-sensitive trust in AI systems, often accepting algorithmic input for technical or data-driven tasks while showing greater reliance on human judgment in social, moral, or ambiguous contexts, with research highlighting that trust in AI is domain-dependent and higher for factual or objective recommendations than for decisions involving subjective or interpersonal evaluation \citep{Afroogh2024, Li2024developing}. This apparent distinction between epistemic trust (focused on factual accuracy) and social trust (centered on subjective guidance) implies engagement of separate cognitive processes \citep{sedlakova2025human, li2023epistemic}. Yet, behavioral choices alone, such as endpoint preferences, reveal little about the underlying dynamics, including the temporal evolution and evidential strength leading to a decision \citep{scharowski2023exploring}.

To address these dynamics, the present study applies the Drift Diffusion Model, a prominent evidence accumulation framework from mathematical psychology and cognitive neuroscience \citep{Ratcliff2008, PerezParra2022a}. The DDM models decisions as noisy evidence accumulation over time toward one of two boundaries, decomposing choice and reaction time data into key parameters: drift rate ($v$), reflecting the speed and direction of evidence accrual (i.e., preference strength); boundary separation ($a$), indicating response caution; starting point ($z$), denoting prior bias; and non-decision time ($t_0$), capturing perceptual and motor delays \citep{Myers2022}. 

Although traditional DDM applications estimate these parameters separately for each participant (or condition), such independent fitting ignores systematic individual differences and trial-to-trial dependencies that are particularly relevant in factual and social decision-making contexts. To overcome these limitations and jointly model both within-subject variability and between-subject heterogeneity while allowing parameters to vary systematically with covariates, we employ a hierarchical sequential sampling model (HSSM; \citealt{Fengler2023}). This Bayesian hierarchical extension of the DDM provides more precise and shrinkage-regularized parameter estimates, improves recovery of individual-level effects, and enables the direct testing of regression relationships between latent cognitive processes and experimental or dispositional predictors.

The HSSM is exceptionally well-suited to adjudicate between competing hypotheses about context-dependent trust. A preference for AI in factual domains could manifest as a higher drift rate (v) toward the AI option, an initial bias in the starting point (z) toward AI, or even a lower decision boundary (a) for AI-related choices, reflecting less deliberation \citep{Lee2023Confidence, CalderTravis2024Bayesian}. The model has already provided profound mechanistic data into related domains of social cognition, including moral decision-making and impression formation \citep{Lin2024Moral, Ledesma2025Moral, Gates2021Preferences}. Furthermore, the HSSM provides a formal framework to describe phenomena such as the “brittleness” of AI trust \citep{geng2025childrens}. In this view, a decision process with a rapid drift rate can reach conclusions quickly but may also exhibit heightened sensitivity to later contradictory evidence, a property that offers a useful way to conceptualize the sharp declines in trust observed after AI errors.

The present study applies a hierarchical Bayesian drift–diffusion model (HSSM) to examine how university students’ trust decisions unfold when choosing between AI and human informants across different domains. This approach allows us to move beyond behavioral outcomes to a process-level account, distinguishing whether context-dependent preferences reflect differences in starting-point bias, decision boundaries, or the rate of evidence accumulation. In doing so, the study addresses a critical gap in the literature: while prior work documents a preference for AI in epistemic contexts and humans in social contexts, the underlying cognitive mechanisms that generate this pattern remain unknown.

In summary, this research addresses an important gap by linking behavioral observations of human–AI trust with a formal, quantitative model of the underlying cognitive processes. By operationalizing trust within an evidence-accumulation framework, we move beyond describing whether and when people trust AI to examining how these decisions unfold in real time. Such process-level evidence is essential for advancing theories of human–AI collaboration and for informing the design of intelligent systems that foster robust, well-calibrated trust. To this end, our investigation is guided by the following research questions:

1. To what extent are university students’ context-dependent trust preferences for AI versus human informants explained by shifts in a priori bias (z), response caution (a), or evidence accumulation rate (v)?

2. What does the magnitude of the difference in drift rates between epistemic and social contexts indicate about the strength and efficiency of the cognitive processes underlying trust in AI versus humans?

3. How do latent cognitive parameters of trust formation, quantified by the HSSM, map onto explicit, self-reported trust judgments, and to what extent does this relationship support the HSSM as a robust framework for studying human-AI interaction?

\section{Methodology}
\subsection{Participants}
A total of 102 undergraduate students from the Psychology and Nursing programs at Universidad de La Sabana (Colombia) participated in the study. The mean age of the participants was 19.318 years ($SD = 5.660$), and the sample comprised 17 men. All participants were native Spanish speakers and were recruited based on accessibility during university classes. 

\subsection{Instruments}
Data on sociodemographic variables (e.g., gender, age, socioeconomic level, access to technology) and prior trust attitudes (e.g., comparative trust in AI and human agents) were collected through a short questionnaire. 

A novel experiment named *Trust in AI: Epistemic and Social Situations* was employed (Fig 1). It consisted of 30 realistic vignettes representing epistemic and social contexts requiring guidance. In each trial, participants were presented with two options: an AI agent (e.g., ChatGPT, Gemini, Claude) consistently displayed on the left, and a human agent consistently displayed on the right. To indicate their preference, participants used a slider that ranged from 0 (complete preference for the AI agent) to 100 (complete preference for the human agent). Intermediate values reflected degrees of trust distribution. 

\begin{figure}[H]
\begin{minipage}{\textwidth}
\raggedright
\textbf{Figure 1} \\[0.5em]
\textit{Experimental interface for the Trust in AI: Epistemic and Social Situations task, illustrating the presentation of epistemic and social vignettes and the continuous response slider ranging from complete AI preference (0) to complete Human preference (100).}
\end{minipage}

\vspace{1em}

\begin{center}
\includegraphics[width=0.8\textwidth]{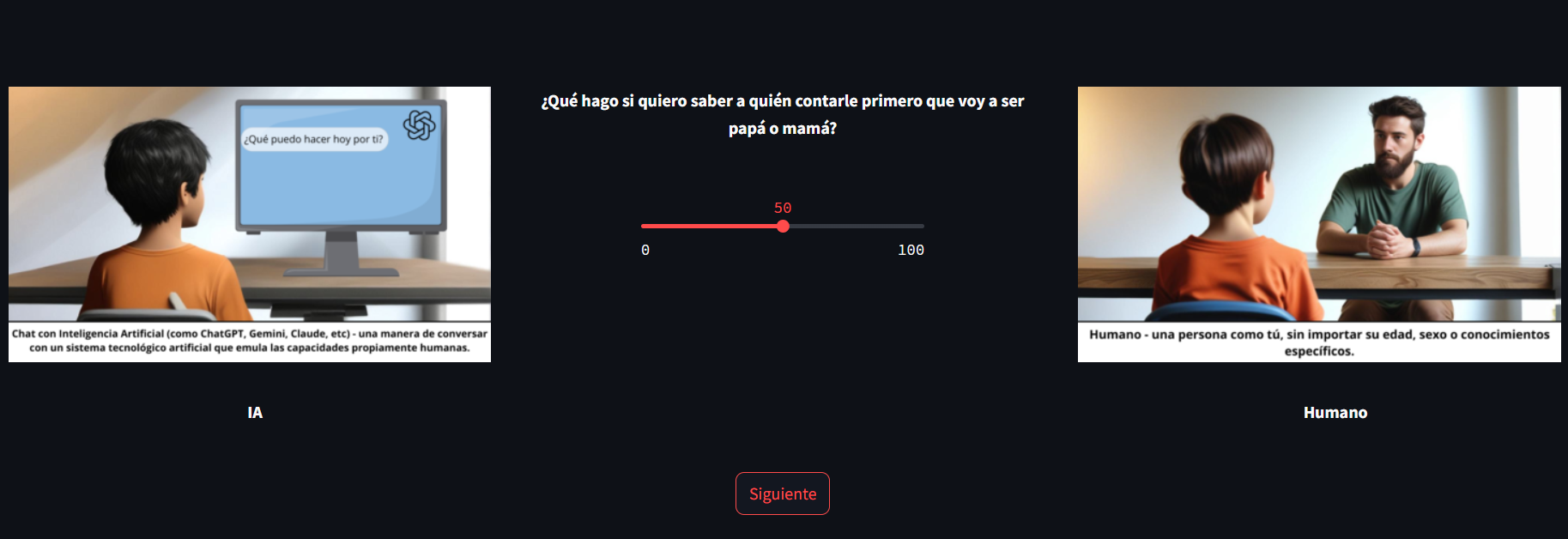}
\end{center}

\vspace{1em}

\begin{minipage}{\textwidth}
\raggedright
\small
\end{minipage}

\label{fig:instrument}
\end{figure}

Neutral visual icons and textual descriptions accompanied both options to ensure clarity and minimize ambiguity (e.g., chatbots like ChatGPT, Gemini, or Claude). This format enabled the collection of three behavioral measures per trial: 1. the selected decision (AI vs. human), 2. the exact slider value indicating trust strength, and 3. the reaction time for the response.

\subsection{Rationale for Methodological Approach}
This study adopted a behavioral decision-making paradigm to infer trust from continuous measures. In line with recent calls within HCI to move beyond exclusive reliance on self-reported trust questionnaires, this approach captures revealed preferences in context-dependent choices. The inclusion of the slider provided a graded measure of trust calibration, while reaction times offered an additional proxy for cognitive effort and decisional conflict \citep{Gulati2024TrustHCI}. 

Together, these measures enabled a more nuanced analysis of trust dynamics. Specifically, they allowed modeling of the latent cognitive processes underlying decision-making through hierarchical drift diffusion models (HSSM). By estimating drift rate ($v$), boundary separation ($a$), non-decision time ($t$), and starting bias ($z$), we were able to quantify how evidence is accumulated when deciding between AI and human agents in different situational contexts. This methodology aligns with recent advances in cognitive modeling of trust and human-AI interaction, offering higher ecological and predictive validity compared to explicit ratings alone.

\subsection{Procedure}
The study was conducted during regular university class sessions. Participants were informed about the purpose of the study (i.e., to explore preferences for guidance in hypothetical decision-making scenarios as a measure of contextual trust) and provided informed consent online. 

After scanning a QR code, participants were directed to a custom-built web application developed with Python’s Streamlit library. The platform first administered the sociodemographic questionnaire, followed by the Trust in AI: Epistemic and Social Situations experiment. Each of the 30 scenarios was presented sequentially, always displaying the two agent options (AI on the left, human on the right) with their respective descriptions and icons. Participants used the slider to indicate their preference, and the system automatically recorded the decision, slider value, and reaction time for each trial. 

Upon completion, participants received a debriefing screen with information about the study’s objectives and additional resources on AI ethics. Data were anonymized immediately, and no incentives were provided beyond integration into course credit.

\subsection{Data Analysis}
Data were analyzed using Hierarchical Sequential Sampling Modeling (HDDM) through the \texttt{HSSM} package \citep{Fengler2025HSSM} in Python. 

A Hierarchical Sequential Sampling Modeling (HSSM) was specified in which the four core parameters, drift rate ($v$), boundary separation ($a$), non-decision time ($t$), and starting point ($z$), were estimated simultaneously at both participant and scenario levels. 

Specifically, the model assumed the following regression structure:

\begin{align}
v_{i[j,k]} &\sim \mathcal{N}\left(\beta_{0}^{v} + u_{j}^{v} + w_{k}^{v}, \sigma_{v}^{2}\right) \\
a_{i[j,k]} &\sim \mathcal{N}^{+}\left(\beta_{0}^{a} + u_{j}^{a} + w_{k}^{a}, \sigma_{a}^{2}\right) \\
t_{i} &\sim \mathcal{N}^{+}\left(\beta_{0}^{t}, \sigma_{t}^{2}\right) \\
z_{i} &\sim \mathrm{Beta}\left(\phi \cdot \eta, (1-\eta)\cdot \phi\right) \quad \text{(reparameterized to } \beta_{0}^{z} = \mathrm{logit}(\eta)\text{)}
\end{align}

where $i$ indexes trials, $j$ indexes participants, $k$ indexes scenarios, $u_{j}^{\cdot} \sim \mathcal{N}(0, \sigma_{\cdot,\text{subject}}^{2})$ and $w_{k}^{\cdot} \sim \mathcal{N}(0, \sigma_{\cdot,\text{situation}}^{2})$ are random effects for subjects and situations, respectively, and $\mathcal{N}^{+}(\cdot)$ denotes a normal distribution truncated at zero (or half-normal prior in the case of $a$ and $t$).

Priors were weakly informative:  
$v \sim \mathcal{N}(0, 1)$,  
$a \sim \mathcal{N}^{+}(\mu=1, \sigma=2)$,  
$t \sim \mathcal{N}^{+}(\sigma=1)$,  
$z \sim \mathrm{Beta}(2, 2)$ (equivalent to a uniform prior on the bias parameter after logit transformation).

Posterior inference was performed using Hamiltonian Monte Carlo (HMC) with the No-U-Turn Sampler (NUTS) implemented in \texttt{NumPyro}. Four Markov chains were run with 5,000 post-warmup draws each (1,000 warmup iterations discarded), yielding 20,000 effective posterior samples. 
Convergence was assessed via $\hat{R} < 1.01$ for all parameters, effective sample sizes (ESS) $> 400$, and visual inspection of trace plots and posterior density overlays.

Model quality and posterior predictive checks were evaluated using ArviZ trace plots, rank plots, posterior predictive distributions, and leave-one-out cross-validation (LOO). 
Forest plots summarized the 95\% highest density intervals (HDIs) of the population-level parameters ($v_{\text{Intercept}}$, $a_{\text{Intercept}}$, $t_{\text{Intercept}}$, $z_{\text{Intercept}}$) and group-level standard deviations. 
To examine contextual effects, posterior distributions of drift rates ($v$) and boundary separation ($a$) were extracted separately for epistemic and social scenarios. 
Finally, 1,000 diffusion trajectories were simulated from the joint posterior (using both stochastic Wiener paths and deterministic mean trajectories) for each condition and visualized to illustrate qualitative differences in evidence accumulation dynamics between epistemic and social guidance-seeking contexts.

To quantify the within-subject association between subjective confidence and evidence-accumulation dynamics, we computed a subject-specific correlation between confidence ratings and the corresponding scenario-level drift-rate estimates derived from the hierarchical sequential sampling model. For each participant, trial-wise confidence values were first rescaled to the interval $[-1,1]$ using $(\text{slider} - 50)/50$. These signed confidence scores were then merged with the posterior means of the effective drift rate $v$ (population intercept plus scenario-specific random offset). Participants with fewer than 30 usable trials or with zero variance in either variable were excluded. For each remaining participant, we computed Pearson’s $r$ between drift rate and signed confidence across all scenarios. Individual $r$ values were Fisher-$z$ transformed and averaged to obtain the group-level estimate, which was then back-transformed for interpretation. Statistical inference was conducted via nonparametric bootstrapping at the subject level ($2{,}000$ resamples), yielding a bias-corrected estimate of the mean correlation and its 95\% confidence interval.

All analyses were conducted in Python 3.12.12 using \texttt{HSSM} v0.2.9, \texttt{NumPyro} v0.19.0, \texttt{ArviZ} v0.22.0, \texttt{Bambi} v0.15.0, and \texttt{ssm-simulators} v0.11.0. 
The complete analysis code, data preprocessing scripts, and posterior samples are openly available at request.

\section{Results}

\subsection{Model Estimation}

The Bayesian Hierarchical Sequential Sampling Model (HSSM) was successfully estimated with random effects for drift rate ($v$) and decision boundary ($a$) by participant and situation. Posterior distributions of the group-level parameters are shown in Figure 2.

\begin{figure}[H]
\begin{minipage}{\textwidth}
\raggedright
\textbf{Figure 2} \\[0.5em]
\textit{Posterior distributions of group-level parameters from the Bayesian Hierarchical Sequential Sampling Model (HSSM), including drift rate ($v$), decision boundary ($a$), non-decision time ($t$), and starting point ($z$).}
\end{minipage}

\vspace{1em}

\begin{center}
\includegraphics[width=0.8\textwidth]{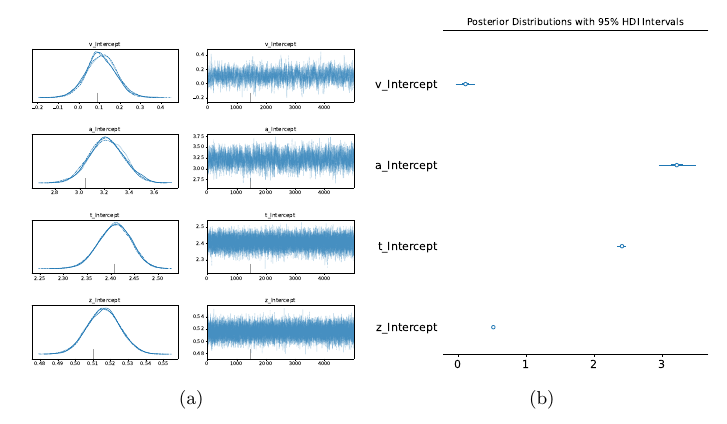}
\end{center}

\vspace{1em}

\begin{minipage}{\textwidth}
\raggedright
\small
\end{minipage}

\label{fig:instrument}
\end{figure}

Posterior summaries indicated that the group-level drift rate ($v$) centered around 0.1, the decision boundary ($a$) around 3.2, the non-decision time ($t$) around 2.4, and the starting point ($z$) around 0.51, as illustrated in Figure 2(a). Figure 2(b) displays the 95\% Highest Density Intervals (HDIs) for each of these posterior parameters.

\subsection{Situation-Level Parameters}

Situation-specific parameter estimates are shown in Figure 2. Panel (a) reports posterior means and 95\% HDIs for drift rate ($v$) across all experimental situations, labeled by condition (epistemic: e, social: s). Panel (b) shows effective decision boundaries ($a$) for the same situations.

\begin{figure}[H]
\begin{minipage}{\textwidth}
\raggedright
\textbf{Figure 3} \\[0.5em]
\textit{Situation-level posterior estimates for the HSSM. Panel (a) shows drift rate ($v$) means and 95\% HDIs for all 30 situations (epistemic vs. social). Panel (b) displays corresponding decision boundary ($a$) estimates.}
\end{minipage}

\vspace{1em}

\begin{center}
\includegraphics[width=0.8\textwidth]{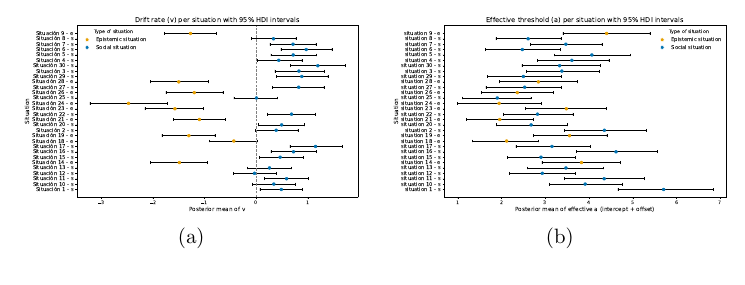}
\end{center}

\vspace{1em}

\begin{minipage}{\textwidth}
\raggedright
\small
\end{minipage}

\label{fig:instrument}
\end{figure}

According to the coding scheme, negative drift rates indicate evidence accumulation favoring AI agents, whereas positive values indicate accumulation favoring Human agents. The situations with the highest negative drift rates were situations 24, 23, 28, and 14 ($v = -2.471, -1.572, -1.498, -1.487$, respectively), all of which corresponded to epistemic situations. In contrast, the situations with the highest positive drift rates were 30, 17, 6, and 29 ($v = 1.194, 1.150, 0.968, 0.884$, respectively), all of which corresponded to social situations.

\subsection{Evidence Accumulation Dynamics}

Figure 3 illustrates representative evidence accumulation trajectories under epistemic and social conditions. Panels (a) and (b) show stochastic simulations with posterior mean parameter values, while panels (c) and (d) present the corresponding deterministic trajectories. 

\begin{figure}[H]
\begin{minipage}{\textwidth}
\raggedright
\textbf{Figure 4} \\[0.5em]
\textit{Representative evidence accumulation trajectories under epistemic and social conditions. Panels (a)–(b) depict stochastic simulations using posterior mean parameters, and panels (c)–(d) present deterministic trajectories illustrating boundary-crossing dynamics.}
\end{minipage}

\vspace{1em}

\begin{center}
\includegraphics[width=0.8\textwidth]{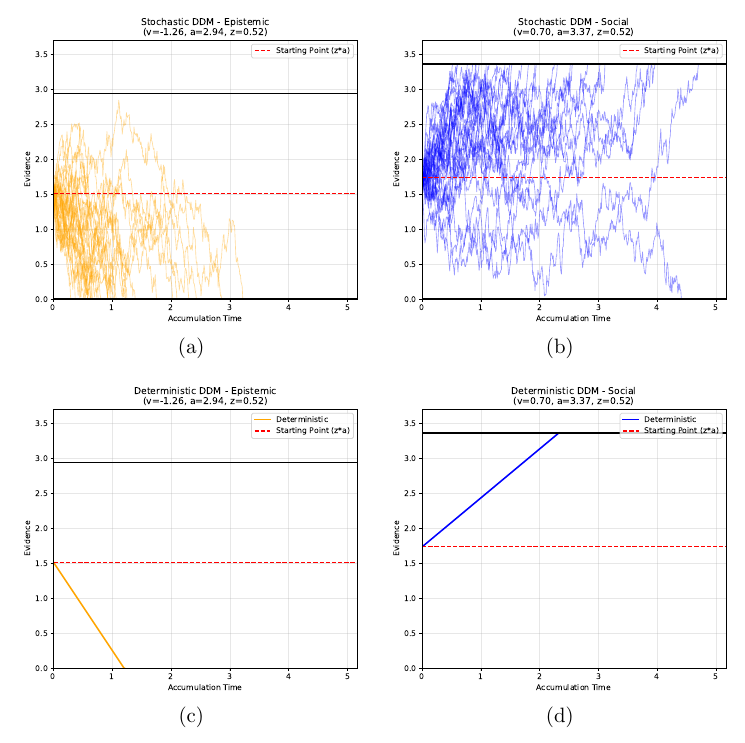}
\end{center}

\vspace{1em}

\begin{minipage}{\textwidth}
\raggedright
\small
\end{minipage}

\label{fig:instrument}
\end{figure}

In the model, the lower decision boundary represented choices for AI agents, whereas the upper boundary represented choices for Human agents. In epistemic conditions, negative drift values produced trajectories converging toward the lower boundary (AI), with mean estimates of $v = -1.26$, $a = 2.94$, and $z = 0.52$. In social conditions, positive drift values produced trajectories converging toward the upper boundary (Human), with mean estimates of $v = 0.70$, $a = 3.37$, and $z = 0.52$.

\subsection{Model-Behavior Correspondence}

Within-subject correlations were computed between signed confidence measured by the position of the slider in every situation, and the posterior mean of the drift rate parameter ($v$) for each of the 30 scenarios. Only participants with valid data and non-zero variance in both variables across all 30 scenarios were included ($N = 93$).

Individual Pearson correlations were positive in 97.8\% of participants (91/93). The Fisher-$z$-transformed correlations were averaged and back-transformed to yield a group-level correlation of $\bar{r} = .736$. A percentile bootstrap procedure (1,000 resamples per participant) produced a 95\% confidence interval of [.699, .766]. The median individual correlation was $r = .762$.

The full distribution of individual within-subject correlations is displayed in Figure 5.

\begin{figure}[H]
\begin{minipage}{\textwidth}
\raggedright
\textbf{Figure 5} \\[0.5em]
\textit{Distribution of within-subject correlations between posterior mean drift rates ($v$) and signed preference ratings (slider values) across all 30 scenarios, summarizing individual-level correspondence between model parameters and behavioral judgments.}
\end{minipage}

\vspace{1em}

\begin{center}
\includegraphics[width=0.8\textwidth]{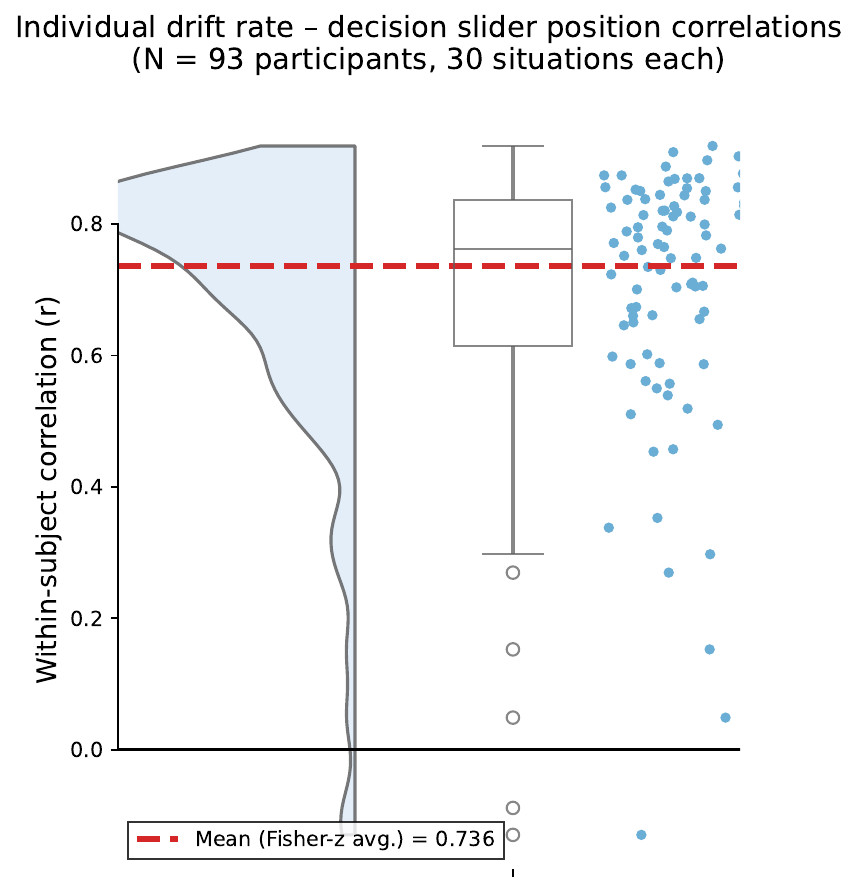}
\end{center}

\vspace{1em}

\begin{minipage}{\textwidth}
\raggedright
\small
\end{minipage}

\label{fig:instrument}
\end{figure}

\section{Discussion}

The present study provides a mechanistic account of context-dependent trust in AI versus human informants among university students, leveraging the Bayesian Hierarchical Model to decompose decision processes into latent cognitive parameters on trust. Our findings reveal that preferences for AI in epistemic domains and humans in social domains are predominantly driven by differences in drift rate ($v$), reflecting varying rates of evidence accumulation toward one option over the other. In epistemic situations, negative drift rates indicate faster and stronger accumulation of evidence favoring AI, while positive drift rates in social situations favor humans. Notably, the starting point ($z$) remained near neutral (approximately 0.52), suggesting minimal a priori bias, and boundary separation ($a$) showed only modest differences, implying similar levels of response caution across contexts. These results directly address our first research question, demonstrating that context-dependent trust shifts arise primarily from adjustments in evidence accumulation efficiency ($v$) rather than initial biases or caution thresholds \citep{Myers2022}.

The magnitude of drift rate differences further illuminates the strength and efficiency of underlying cognitive processes in trust, answering our second research question. In epistemic contexts, the larger absolute drift rates (mean $v = -1.26$) compared to social contexts (mean $v = 0.70$) suggest that students process evidence more decisively and rapidly when favoring AI for factual guidance. This asymmetry may reflect the perceived reliability of AI in objective domains, where algorithmic precision is valued, versus the nuanced, empathetic qualities attributed to humans in interpersonal matters \citep{sahebi2025ai, li2023epistemic, hoehl, Stower2024}. Such patterns align with dual-process theories, where System 1 (fast or mindless processing) intuitions might drive quick AI preferences in low-ambiguity epistemic tasks, while System 2 (slower and analytical responses) deliberation supports human preferences in socially complex scenarios \citep{chiriatti2025system0, zhou2024mindlessly, okonkwo2025explainable}.

To enrich this interpretation, the concept of epistemic vigilance provides a useful lens. Epistemic vigilance refers to the cognitive mechanisms humans employ to evaluate the reliability of communicated information and avoid misinformation \citep{sperber2010epistemic}. In contexts of selective trust, where individuals provisionally accept a source's claims without immediate verification, relying instead on past reliability or contextual cues, such vigilance becomes central in calibrating selection \citep{levy2022trust}. Applied to human–AI interactions, our Drift Diffusion Model results suggest that epistemic vigilance modulates drift rates in a domain-sensitive manner. In epistemic contexts, faster accumulation toward AI may reflect reduced vigilance, as participants defer trust to AI's perceived objectivity, potentially weakening critical evaluation \citep{jose2025digital}. Conversely, in social contexts, slower or positive drifts toward humans appear consistent with heightened vigilance toward AI, where anthropomorphic cues or perceived limitations trigger skepticism about its suitability for subjective matters, prompting greater deference to human informants \citep{jose2025epistemic}. This perspective implies that selective trust in AI is not uniform but contextually gated by vigilance mechanisms, which may help explain the “brittleness” of AI trust: rapid initial deference in factual domains becomes fragile and liable to collapse when errors reactivate vigilance \citep{bartsch2024trust}.

These findings also align with the Extended Epistemic Vigilance Framework (EEVF), which expands traditional models by incorporating evaluation of the receiver alongside the source and the content of the message \citep{bielik2025developing}. In our study, the modest differences observed in boundary separation ($a$) may reflect receiver-oriented vigilance, wherein students maintain a relatively stable level of caution across domains but modulate evidence accumulation based on self-perceived biases, contextual expectations, or metacognitive awareness. This interpretation is consistent with EEVF's emphasis on the role of cognitive biases and socioemotional factors shaping the receiver's evaluation process.

Furthermore, our results connect to research on selective trust, particularly in technological contexts. Young children trust accurate robots for novel information \citep{Stower2024}, and selectively endorse accurate robots over inaccurate ones when perceiving them as agentic \citep{brink2020robot}. In our adult sample, the domain-specific drift rates mirror this selectivity: stronger negative drifts for AI in epistemic tasks indicate selective trust in AI's accuracy for factual matters, akin to children's preferences for accurate technological informants \citep{geng2025childrens}. However, unlike younger children who may trust inaccurate humans more than inaccurate robots \citep{tan2023younger}, our students showed context-dependent selectivity, distrusting AI in social domains. This developmental progression suggests that epistemic vigilance matures to encompass both epistemic reliability and social agency, with adults applying more nuanced selective trust in human-AI interactions. By linking bayesian parameters to epistemic vigilance and selective trust, our results highlight how selective trust emerges from dynamic evidence processing, offering a nuanced view of when and why users might over-rely on AI without sufficient scrutiny.

Moreover, drift rate and explicit trust judgments were positively related within subjects (Fisher-z-averaged r = .736, 95\% CI [.699, .766]), with 97.8\% of individual correlations in the expected direction (N = 93). These results confirm that hierarchically estimated drift rate serves as a valid quantitative index of moment-to-moment trust, addressing our third research question. This convergence supports the model as a robust framework for studying human-AI interactions, capturing not only endpoint choices but also the temporal unfolding of trust formation. These results extend prior behavioral observations of domain-specific trust \citep{osborne2025machine} by elucidating the computational mechanisms, such as differential sensitivity to evidence, that generate brittleness in AI trust after errors \citep{geng2025childrens}.

Our results have broader implications for understanding trust dynamics in human-AI interaction. By demonstrating that context-dependent trust arises from differential evidence accumulation (drift rates) rather than a priori biases, the study reveals dissociable mechanisms underlying trust in AI for factual versus social guidance. This mechanistic distinction, where epistemic domains show faster evidence accrual toward AI while social domains favor humans, suggests that interventions to calibrate AI trust should target evidence processing efficiency rather than merely addressing initial biases or response caution. 

However, these findings also highlight epistemological and governance challenges in human-AI interaction. Dependence on epistemically opaque AI systems may erode established knowledge practices that rely on accountability and transparency among human agents \citep{Koskinen2024}. Our results suggest this erosion operates through rapid, low-vigilance evidence accumulation in epistemic domains, precisely where AI opacity is most problematic. The asymmetry in drift rates indicates that users may defer trust to AI in factual contexts without engaging sufficient critical scrutiny, creating a need for "watchful trust" rather than blind reliance \citep{Lahusen2024}. 

Building on this notion, fostering watchful trust requires not only user-level vigilance but also AI systems intentionally designed to elicit and sustain such vigilance. There is a need for cross-disciplinary work to clarify where the boundaries of trust lie and to identify the circumstances that enable the emergence and sustained presence of a watchful strategies, along with the processes that support the effective functioning of such vigilance. \citep{Lahusen2024}. Our findings show why such mechanisms are needed: when evidence accumulation is too rapid or unreflective, users lack the cognitive triggers required to activate epistemic vigilance, giving rise to trust that is excessive and easily destabilized. 

Rather than reflecting distrust, this instability stems from trust granted without calibration, confidence that is accepted by default and thus vulnerable to disruption. Designing for watchful trust therefore requires embedding cues, friction points, and uncertainty signals that interrupt automatic acceptance and prompt critical appraisal. Through calibrated transparency, context-sensitive explanations, or epistemic checkpoints, AI systems can scaffold users’ ability to monitor and adjust their trust dynamically, enhancing stability without reducing efficiency.

At a broader institutional level, this tendency toward unexamined reliance highlights the need for governance strategies that sustain epistemic vigilance rather than assuming that technical trustworthiness alone is sufficient. Regulatory instruments, such as mandated uncertainty disclosures or human oversight in high-stakes contexts, can help maintain an appropriate level of reflective engagement, tempering over-reliance while preserving the efficiency that makes AI appealing for epistemic tasks. Aligning cognitive requirements with institutional safeguards thus suggests that trustworthy AI depends not only on algorithmic performance but also on socio-technical arrangements that reinforce calibrated and vigilant trust.

Despite these contributions, limitations must be acknowledged. The sample was drawn from a single Colombian university, potentially limiting generalizability to diverse cultural or age groups. The hypothetical scenarios, while ecologically inspired, may not fully capture real-world decision dynamics, where factors like AI errors or interpersonal rapport could influence parameters.

Future research should extend this paradigm to longitudinal designs, tracking how experience with AI modulates HSSM parameters over time, or to neuroimaging studies linking drift rates to neural correlates of trust. Cross-cultural comparisons could also reveal how societal norms shape evidence accumulation in human-AI choices.

In conclusion, by quantifying trust dynamics through HSSM, our study moves beyond descriptive approaches and demonstrates how evidence is accumulated when individuals decide whether to trust AI. Our process-level perspective reveals that context-dependent preferences arise from differential evidence accumulation, offering a more mechanistic understanding of selective trust in AI. These results provide a foundation for more nuanced theories of human–AI collaboration and contribute to the development of more trustworthy intelligent systems.

\section{Declaration of generative AI and AI-assisted technologies in the writing process}

During the preparation of this work, the author(s) used ChatGPT, an AI language model developed by OpenAI, to support the writing process by enhancing the clarity and comprehension of the text. After using this tool, the author(s) thoroughly reviewed and edited the content as needed, taking full responsibility for the final version of the published article.

\bibliography{references}

\appendix

\section*{Appendix A: Situations in Spanish used in experiment}

\begin{itemize}
    \item ¿Qué hago si quiero saber a quien contarle primero que voy a ser papá o mamá?  
    \item ¿Qué hago si no sé cómo avisarles a mis padres que tengo malas calificaciones?  
    \item ¿Qué hago si quiero cuidar bien a un bebé?  
    \item ¿Qué hago cuando necesito saber si está bien o no hacer trampa en un juego o negocio?  
    \item ¿Qué hago si veo a alguien copiando en un examen y no estoy seguro si debo avisarle al profesor?  
    \item ¿Qué hago si alguien se burla de mí?  
    \item ¿Qué hago si me siento triste y quiero saber si debería hablar con alguien de eso?  
    \item ¿Qué hago para saber si mis ideas de alguien son correctas o no?  
    \item ¿Qué hago si quiero saber si los gatos ven mejor que los perros en la oscuridad?  
    \item ¿Qué hago si no puedo dejar de hacer algo que creo me hace daño?  
    \item ¿Qué hago cuando necesito saber si es bueno pedir un consejo al momento de tomar decisiones importantes?  
    \item ¿Qué hago si quiero saber si es bueno o malo lo que pide una religión?  
    \item ¿Qué hago para saber qué hacer si alguien me ha golpeado y me duele?  
    \item ¿Qué hago si quiero saber cuál es el mejor jugador de futbol del mundo?  
    \item ¿Qué hago si quiero saber si soy o no físicamente atractivo/a?  
    \item ¿Qué hago si quiero saber si debo decir que algún familiar o alguien cercano se comporta mal conmigo?  
    \item ¿Qué hago si me siento mal porque creo que le he hecho daño a alguien?  
    \item ¿Qué hago si quiero sembrar un árbol de mango en casa?  
    \item ¿Qué hago si quiero saber cuál es la mejor época para viajar a otro país?  
    \item ¿Qué hago si necesito saber si estar enamorado es bueno o no?  
    \item ¿Qué hago si quiero saber cuál es la mejor comida del mundo?  
    \item ¿Qué hago si me invitan a consumir drogas o alcohol?  
    \item ¿Qué hago si quiero saber el año exacto en que inventaron el bombillo?  
    \item ¿Qué hago si quiero saber cuál es el nombre exacto de Shakira?  
    \item ¿Qué hago si quiero saber si Dios existe o no?  
    \item ¿Qué hago si quiero preparar un pastel de chocolate?  
    \item ¿Qué hago si quiero vengarme de alguien?  
    \item ¿Qué hago si quiero saber cuál es el mejor momento del año para ir a la playa?  
    \item ¿Qué hago si quiero saber qué sentido tiene mi vida?  
    \item ¿Qué hago si me siento alegre y quiero saber si debería hablar con alguien de eso?  
\end{itemize}

\newpage

\section*{Appendix B: Situations translated in English}

\begin{itemize}
    \item What should I do if I want to know who to tell first that I am going to be a mom or dad?  
    \item What should I do if I don't know how to tell my parents that I have bad grades?  
    \item What should I do if I want to know how to take good care of a baby?  
    \item What should I do when I need to know if it's okay or not to cheat in a game or business?  
    \item What should I do if I see someone cheating on a test and I'm not sure if I should tell the teacher?  
    \item What should I do if someone makes fun of me?  
    \item What should I do if I feel sad and want to know if I should talk to someone about it?  
    \item What should I do to know if my opinions about someone are right or not?  
    \item What should I do if I want to know if cats see better than dogs in the dark?  
    \item What should I do if I can't stop doing something that I think is hurting me?  
    \item What should I do when I need to know if it's good to ask for advice when making important decisions?  
    \item What should I do if I want to know if what a religion asks for is good or bad?  
    \item What should I do to know what to do if someone hit me and it hurts?  
    \item What should I do if I want to know who the best soccer player in the world is?  
    \item What should I do if I want to know if I am physically attractive or not?  
    \item What should I do if I want to know if I should say that a relative or someone close to me is misbehaving with me?  
    \item What should I do if I feel bad because I think I hurt someone?  
    \item What should I do if I want to plant a mango tree at home?  
    \item What should I do if I want to know the best time of year to travel to another country?  
    \item What should I do if I need to know if being in love is good or not?  
    \item What should I do if I want to know what the best food in the world is?  
    \item What should I do if someone invites me to use drugs or alcohol?  
    \item What should I do if I want to know the exact year the light bulb was invented?  
    \item What should I do if I want to know Shakira's exact name?  
    \item What should I do if I want to know if God exists or not?  
    \item What should I do if I want to make a chocolate cake?  
    \item What should I do if I want to get revenge on someone?  
    \item What should I do if I want to know the best time of the year to go to the beach?  
    \item What should I do if I want to know the meaning of my life?  
    \item What should I do if I feel happy and want to know if I should talk to someone about it?  
\end{itemize}

\end{document}